\UseRawInputEncoding
\documentclass[preprint2]{aastex631}

\begin{document}

\title{{\sl NEOWISE-R} Caught the Luminous SN 2023ixf in Messier 101}

\correspondingauthor{Schuyler Van Dyk}
\email{vandyk@ipac.caltech.edu}

\author[0000-0001-9038-9950]{Schuyler D.~Van Dyk}
\affiliation{Caltech/IPAC, Mailcode 100-22, Pasadena, CA 91125, USA}

\author[0000-0003-4610-1117]{Tam\'as Szalai}
\affiliation{Department of Experimental Physics, Institute of Physics, University of Szeged, D{\'o}m t{\'e}r 9, Szeged, 6720, Hungary}
\affiliation{MTA-ELTE Lend\"ulet "Momentum" Milky Way Research Group, Hungary}

\author[0000-0002-0077-2305]{Roc M.~Cutri}
\affiliation{Caltech/IPAC, Mailcode 100-22, Pasadena, CA 91125, USA}

\author[0000-0003-4269-260X]{J.~Davy Kirkpatrick}
\affiliation{Caltech/IPAC, Mailcode 100-22, Pasadena, CA 91125, USA}

\author[0000-0003-4072-169X]{Carl J.~Grillmair}
\affiliation{Caltech/IPAC, Mailcode 314-6, Pasadena, CA 91125, USA}

\author[0000-0001-9309-0102]{Sergio B.~Fajardo-Acosta}
\affiliation{Caltech/IPAC, Mailcode 100-22, Pasadena, CA 91125, USA}

\author[0000-0003-2638-720X]{Joseph R.~Masiero}
\affiliation{Caltech/IPAC, Mailcode 100-22, Pasadena, CA 91125, USA}

\author[0000-0002-7578-3885]{Amy K.~Mainzer}
\affiliation{University of Arizona, 1629 E.~University Boulevard, Tucson, AZ 85721, USA}
\affiliation{Department of Earth, Planetary, and Space Sciences, The University of California, Los Angeles, 595 Charles E.~Young Drive East, Los Angeles, CA 90095, USA}

\author[0000-0001-5072-4574]{Christopher R.~Gelino}
\affiliation{Caltech/IPAC, Mailcode 100-22, Pasadena, CA 91125, USA}

\author[0000-0001-8764-7832]{J\'ozsef Vink\'o}
\affiliation{Department of Experimental Physics, Institute of Physics, University of Szeged, D{\'o}m t{\'e}r 9, Szeged, 6720, Hungary}
\affiliation{HUN-REN CSFK Konkoly Observatory,  Konkoly Th. M. {\'u}t 15-17, Budapest, 1121, Hungary}
\affiliation{CSFK, MTA Centre of Excellence, Konkoly Thege Mikl{\'o}s {\'u}t 15-17, Budapest, 1121, Hungary}
\affiliation{ELTE E{\"o}tv{\"o}s Lor{\'a}nd University, Institute of Physics and Astronomy, P{\'a}zm{\'a}ny P{\'e}ter s{\'e}t{\'a}ny 1/A, Budapest, 1117, Hungary}

\author[0000-0001-5203-434X]{Andr\'as P\'eter Jo\'o}
\affiliation{HUN-REN CSFK Konkoly Observatory,  Konkoly Th. M. {\'u}t 15-17, Budapest, 1121, Hungary}
\affiliation{CSFK, MTA Centre of Excellence, Konkoly Thege Mikl{\'o}s {\'u}t 15-17, Budapest, 1121, Hungary}
\affiliation{ELTE E{\"o}tv{\"o}s Lor{\'a}nd University, Institute of Physics and Astronomy, P{\'a}zm{\'a}ny P{\'e}ter s{\'e}t{\'a}ny 1/A, Budapest, 1117, Hungary}

\author[0000-0001-5449-2467]{Andr\'as P\'al}
\affiliation{HUN-REN CSFK Konkoly Observatory,  Konkoly Th. M. {\'u}t 15-17, Budapest, 1121, Hungary}
\affiliation{CSFK, MTA Centre of Excellence, Konkoly Thege Mikl{\'o}s {\'u}t 15-17, Budapest, 1121, Hungary}

\author[0000-0002-8770-6764]{R\'eka K\"onyves-T\'oth}
\affiliation{HUN-REN CSFK Konkoly Observatory,  Konkoly Th. M. {\'u}t 15-17, Budapest, 1121, Hungary}
\affiliation{CSFK, MTA Centre of Excellence, Konkoly Thege Mikl{\'o}s {\'u}t 15-17, Budapest, 1121, Hungary}
\affiliation{Department of Experimental Physics, Institute of Physics, University of Szeged, D{\'o}m t{\'e}r 9, Szeged, 6720, Hungary}
\affiliation{ELTE Eötvös Loránd University, Gothard Astrophysical Observatory, Szombathely, Hungary}

\author{Levente Kriskovics}
\affiliation{HUN-REN CSFK Konkoly Observatory,  Konkoly Th. M. {\'u}t 15-17, Budapest, 1121, Hungary}
\affiliation{CSFK, MTA Centre of Excellence, Konkoly Thege Mikl{\'o}s {\'u}t 15-17, Budapest, 1121, Hungary}

\author[0000-0002-1698-605X]{R\'obert Szak\'ats}
\affiliation{HUN-REN CSFK Konkoly Observatory,  Konkoly Th. M. {\'u}t 15-17, Budapest, 1121, Hungary}
\affiliation{CSFK, MTA Centre of Excellence, Konkoly Thege Mikl{\'o}s {\'u}t 15-17, Budapest, 1121, Hungary}

\author[0000-0002-6471-8607]{Kriszti\'an Vida}
\affiliation{HUN-REN CSFK Konkoly Observatory,  Konkoly Th. M. {\'u}t 15-17, Budapest, 1121, Hungary}
\affiliation{CSFK, MTA Centre of Excellence, Konkoly Thege Mikl{\'o}s {\'u}t 15-17, Budapest, 1121, Hungary}

\author[0000-0002-2636-6508]{WeiKang Zheng}
\affiliation{Department of Astronomy, University of California, Berkeley, CA 94720-3411, USA}

\author[0000-0001-5955-2502]{Thomas G.~Brink}
\affiliation{Department of Astronomy, University of California, Berkeley, CA 94720-3411, USA}

\author[0000-0003-3460-0103]{Alexei V.~Filippenko}
\affiliation{Department of Astronomy, University of California, Berkeley, CA 94720-3411, USA}

\begin{abstract}
The reactivated {\sl Near-Earth Object Wide-field Infrared Survey Explorer\/} ({\sl NEOWISE-R}) serendipitously caught the Type II supernova SN 2023ixf in Messier 101 on the rise, starting day 3.6 through day 10.9, and on the late-time decline from days 211 through 213 and days 370 through 372. We have considered these mid-infrared (mid-IR) data together with observations from the ultraviolet (UV) through the near-IR, when possible. At day 3.6 we approximated the optical emission with a hot, $\sim 26,630$ K blackbody, with a notable UV excess likely from strong SN shock interaction with circumstellar matter (CSM). In the IR, however, a clear excess is also obvious, and we fit it with a cooler, $\sim 1,620$ K blackbody with radius of $\sim 2.6 \times 10^{15}$ cm, consistent with dust in the progenitor's circumstellar shell likely heated by the UV emission from the CSM interaction. On day 10.8, the light detected was consistent with SN ejecta-dominated emission. At late times we also observed a clear {\sl NEOWISE-R\/} excess, which could arise either from newly formed dust in the inner ejecta or in the contact discontinuity between the forward and reverse shocks, or from more distant pre-existing dust grains in the SN environment. Furthermore, the large 4.6 $\mu$m excess at late times can also be explained by the emergence of the carbon monoxide 1--0 vibrational band. SN 2023ixf is the best-observed SN IIP in the mid-IR during the first several days after explosion and one of the most luminous such SNe ever seen.
\end{abstract}


\section{Introduction}\label{sec:intro}

Supernova (SN) explosions are among the most powerful events in the Universe. They serve as unique cosmic laboratories for studying processes in extreme physical conditions and chemical feedback into the interstellar and intergalactic media. Core-collapse supernovae (CCSNe), consequences of the gravitational collapse of iron cores of massive ($\gtrsim 8\ M_{\odot}$) stars, have been considered as possible sources of cosmic dust at high redshifts for over $\sim 50$~yr \citep[e.g.,][]{cernuschi67, hoyle70,dwek07}. Observed dust in CCSNe may form either in the (unshocked) ejecta or in a cold dense shell (CDS) across the contact discontinuity between the shocked circumstellar matter (CSM) and shocked ejecta. A late-time mid-infrared (mid-IR) excess may emerge from either newly formed or heated pre-existing dust grains. In the shocked CSM, heating can be collisional, and grains in the more distant, unshocked CSM are assumed to be radiatively heated by the peak SN luminosity or by energetic photons generated during CSM interaction (see, e.g., \citealt{gall11} for a review).

Observed properties and classifications of CCSNe, from H-rich to H-free spectra (IIP, IIL, IIb, Ib/c; see \citealt{filippenko97} and \citealt{GalYam2017} for reviews), depend mainly on the degree of pre-explosion mass loss from the progenitor star (and/or its companion in a binary system). H-rich Type II-plateau (IIP) SNe, characterized by their $\sim100$-day optical plateau-like light curves, are the predominant CCSN subclass \citep[corresponding to $\sim 55$\% of all CCSNe;][]{Perley_2020}. Direct evidence exists that these SNe arise from stars in the red supergiant (RSG) phase, with the star's massive hydrogen envelope remaining relatively intact at explosion \citep{Smartt2009,Smartt2015,VanDyk2017}. SNe IIP are known to form new dust in their ejecta (see \citealt{gall11} and references therein). Based on recent model calculations, expanding SN~IIP ejecta succeed in condensing sufficient quantities (0.05--$1.0\ M_{\odot}$) of dust. Some of these models propose slow and steady dust growth over several thousand days \citep[e.g.,][]{gall14,wesson15}, while others suggest a more rapid dust growth \citep{dwek19,sarangi22}. These theoretical expectations are also in agreement with the far-infrared/submillimeter detection of a large amount of cold ($\lesssim 50$~K) dust in hundreds-of-years-old Galactic SN remnants, such as Cas~A \citep{barlow10,sibthorpe10,arendt14} and the Crab Nebula \citep{gomez12,temim13,delooze19}, as well as the nearby ($\sim 50$~kpc) and famous SN~1987A \citep{matsuura11,matsuura19,indebetouw14}. Very recently, the {\sl James Webb Space Telescope\/} ({\sl JWST}) has offered a new opportunity to study the late phases of cool ($\sim 100$--200~K) dust in extragalactic SNe and has already led to the detection of a significant amount ($\gtrsim10^{-3}\ M_{\odot}$) of dust in Type IIP SNe~2004et and 2017eaw \citep{Shahbandeh_2023}, and the Type IIL SN~1980K \citep{Zsiros_2024}. 

At the same time, only a handful of nearby young ($\lesssim$5~yr) SNe II (primarily IIP) show direct observational evidence for dust condensation, and these examples have all yielded two-to-three orders of magnitude less dust ($\sim 10^{-5}$--10$^{-3}\ M_{\odot}$) than predicted by the models. Most of these observations, however, were carried out in the wavelength range 3--$5\ \mu$m, and thus have been limited to just the warmer ($\gtrsim 500$~K) dust grains. In the last quarter century, the primary source of mid-infrared (mid-IR) data on SNe was NASA's {\sl Spitzer Space Telescope}, which resulted in valuable data during both its cryogenic (2003--2009) and post-cryogenic (2009--2020) missions. Except for several single-object studies --- e.g., SN~1987A \citep{bouchet06,dwek10}, SN~1993J \citep{zsiros22}, SN~1995N \citep{vandyk13}, SN~2003gd \citep{sugerman06,meikle07}, SN~2004dj \citep{szalai11,meikle11}, SN~2004et \citep{kotak09,fabbri11}, SN~2005af \citep{kotak06,szalai13}, SN~2005ip \citep{fox10}, SN~2007it \citep{andrews11b}, and SN~2007od \citep{andrews10} --- most of these {\it Spitzer\/} SN data were collected in the post-cryogenic phase (at 3.6 and 4.5 $\mu$m). Studies during this phase included either targeted surveys, such as the SPIRITS project \citep[SPitzer InfraRed Intensive Transients Survey, a systematic study of transients in nearby galaxies; see][]{tinyanont16,kasliwal17,jencson19} and work focused on interacting SNe \citep{fox11,fox13,szalai21}, or archival images for which the SNe were not the primary target \citep{szalai19a}. The latter work, including data from targeted surveys, presents the most extensive analysis of mid-IR SN observations to date, including $\sim 120$ positively detected objects from $\sim 1100$ SN sites imaged by {\sl Spitzer}.

Another very important tool for detecting early-time mid-IR radiation from SNe has been the {\sl Wide-field Infrared Survey Explorer\/} ({\sl WISE}, both cryogenic and post-cryogenic, 2009--2011; \citealt{Wright2010}). The post-cryogenic {\sl WISE\/} mission was reactivated in 2013 and has been monitoring the sky at 3.4 and 4.6 $\mu$m ever since, as the {\sl Near-Earth Object Wide-field Infrared Survey Explorer Reactivation\/} ({\sl NEOWISE-R}, or {\sl NEOWISE\/} for short; \citealt{Mainzer2011,Mainzer2014}). While the original aim of the reactivated mission is mainly characterization of known Solar System objects, its database also serves as a valuable source of information on a rich variety of transient objects, such as cataclysmic variables, active galactic nuclei, tidal disruption events, and SNe (e.g., \citealt{Kokubo_2019,Tartaglia_2020,Sun_2022,Moran2023,Wang2024}).

We as a community have been incredibly fortunate to have the recent, nearby SN 2023ixf occur in Messier 101 (M101; NGC 5457). Its proximity and brightness have led to many investigators training various facilities at a range of wavelengths at the event, which has exhibited a number of fascinating properties. The SN was discovered by \citet{Itagaki2023} on 2023 May 19.73 (UTC dates are used throughout this paper) and classified as an SN II by \citet{Perley2023} within hours of discovery. It was evident immediately that the optical spectrum was dominated by ``flash'' emission features indicative of interaction of the SN shock with pre-existing CSM \citep[e.g.,][]{Jacobson-Galan2023,Bostroem2023,Hiramatsu2023,Teja2023}. The SN's light curves provided similar indications \citep[e.g.,][]{Hosseinzadeh2023,Hiramatsu2023,Martinez2024}. \citet{Zimmerman2024}, from an analysis of early-time {\sl Hubble Space Telescope\/} ({\sl HST}) ultraviolet (UV) spectroscopy of the SN, constrained the CSM to be dense and confined, with $\sim 10^{-12}$ g cm$^{-3}$ at $\lesssim 2 \times 10^{14}$ cm; they concluded that this dense CSM immediate to the progenitor prolonged the SN shock breakout by $\sim 3$ d. Other indications of initial and longer-term CSM interaction for SN 2023ixf come from observations at X-ray \citep{Grefenstette2023,Chandra2024} and radio \citep{Berger2023} wavelengths.

A progenitor candidate was directly identified in archival {\sl HST\/}, {\sl Spitzer\/}, and ground-based near-IR data \citep[e.g.,][]{Pledger2023,Kilpatricketal2023,Jencson2023,Soraisam2023,VanDyk2024}. These unprecedented data were plentiful enough that the star was shown in astonishing detail to be a long-period variable, similar to what we expect for many RSGs \citep{Jencson2023,Soraisam2023}. Additionally, modeling of the star's spectral energy distribution (SED), e.g., by \citet{VanDyk2024} revealed it to be quite dusty and luminous, and implied that the star was surrounded by a dusty silicate-rich shell with an inner radius of $\approx 10$ times the star's radius, or $\approx 10^{15}$ cm.

Near-IR studies of SN 2023ixf have already been conducted and published \citep[][]{Yamanaka2023,Teja2023}, and others will likely emerge. The SN has already been observed with the {\sl JWST}, and those results are pending. Here we describe and analyze observations by {\sl NEOWISE}, which serendipitously caught SN 2023ixf in the act between $\sim 3$ days and $\sim 372$ days in age. This is among the earliest that an SN has been detected in the mid-IR (the peculiar Type IIP SN~2009js was caught by {\sl Spitzer\/} two days after discovery, however, its explosion epoch is rather uncertain; \citealt{Gandhi_2013,szalai19a}).

Following \citet{Hosseinzadeh2023}, we have adopted 2023 May 18 18:00 UTC (MJD 60082.75) as the explosion epoch. We assume throughout a distance to M101 of $6.85 \pm 0.13$ Mpc \citep{Riess2022}. 

\section{Observations} \label{sec:observations}

\subsection{{\sl NEOWISE}}\label{sec:neowise}

{\sl NEOWISE\/} observed the SN site pre-explosion, as part of routine operations, 152 times between 2013 December 18.26 and 2022 December 18.99 UT (MJD 56644.2618 and 59931.9958, respectively). The first pre-SN pair of single exposures occurred 3438.49 d prior to explosion, while the last was 150.75 d pre-SN. The progenitor candidate was not detected in any of these observations \citep[e.g.,][]{Hiramatsu2023,Jencson2023,Soraisam2023,VanDyk2024}. See also Section~\ref{sec:appendix} of this paper. The lack of detection tends to rule out luminous eruptions or outbursts from the progenitor candidate during that time period prior to explosion, as discussed in the above studies. (The star was also not detected by {\sl WISE\/} and the post-cryogenic {\sl NEOWISE} prior to reactivation, between 2009 and 2011; see \citealt{VanDyk2024}.)

\begin{figure*}[ht!]
\plotone{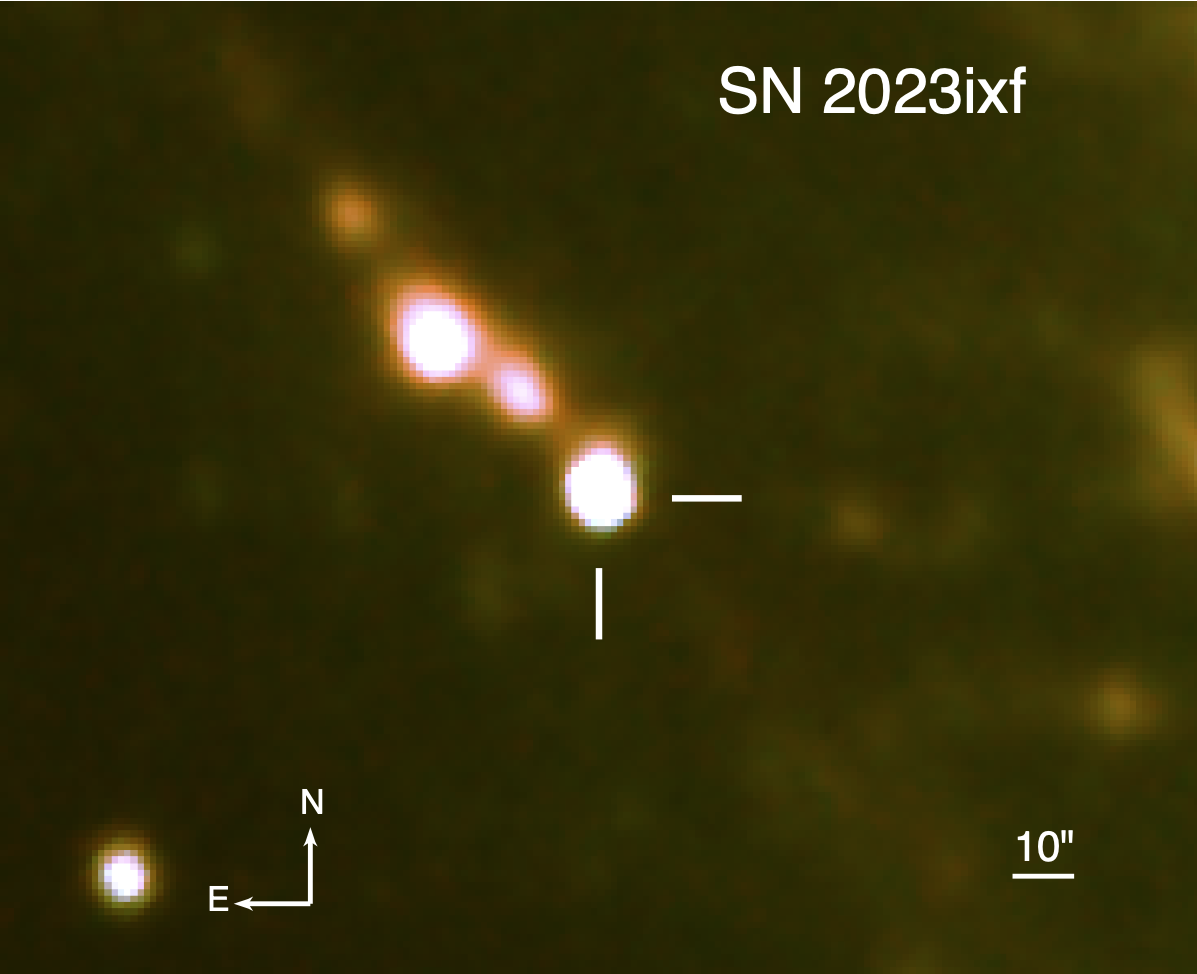}
\caption{Color-composite image of the {\sl NEOWISE-R\/} detection of SN 2023ixf in M101 at 3.4 and 4.6 $\mu$m (bands W1 and W2, respectively) in the day 3.361--4.995 range combined (see Table~\ref{tab:neowise_obs}). The SN is indicated with tick marks. The bright emission to the northeast of the SN is from the giant H~{\sc ii} region complex NGC~5461.
\label{fig:neowise-image}}
\end{figure*}

\begin{figure*}[ht!]
\plottwo{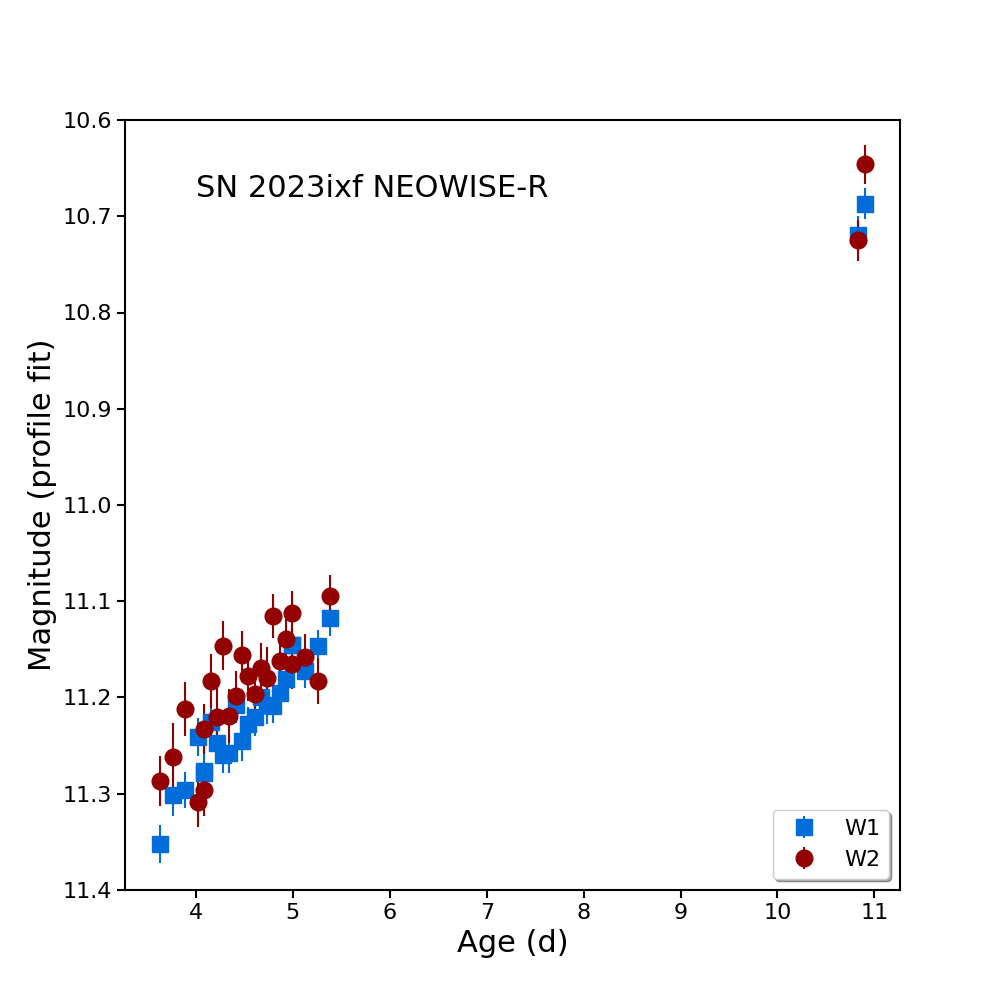}{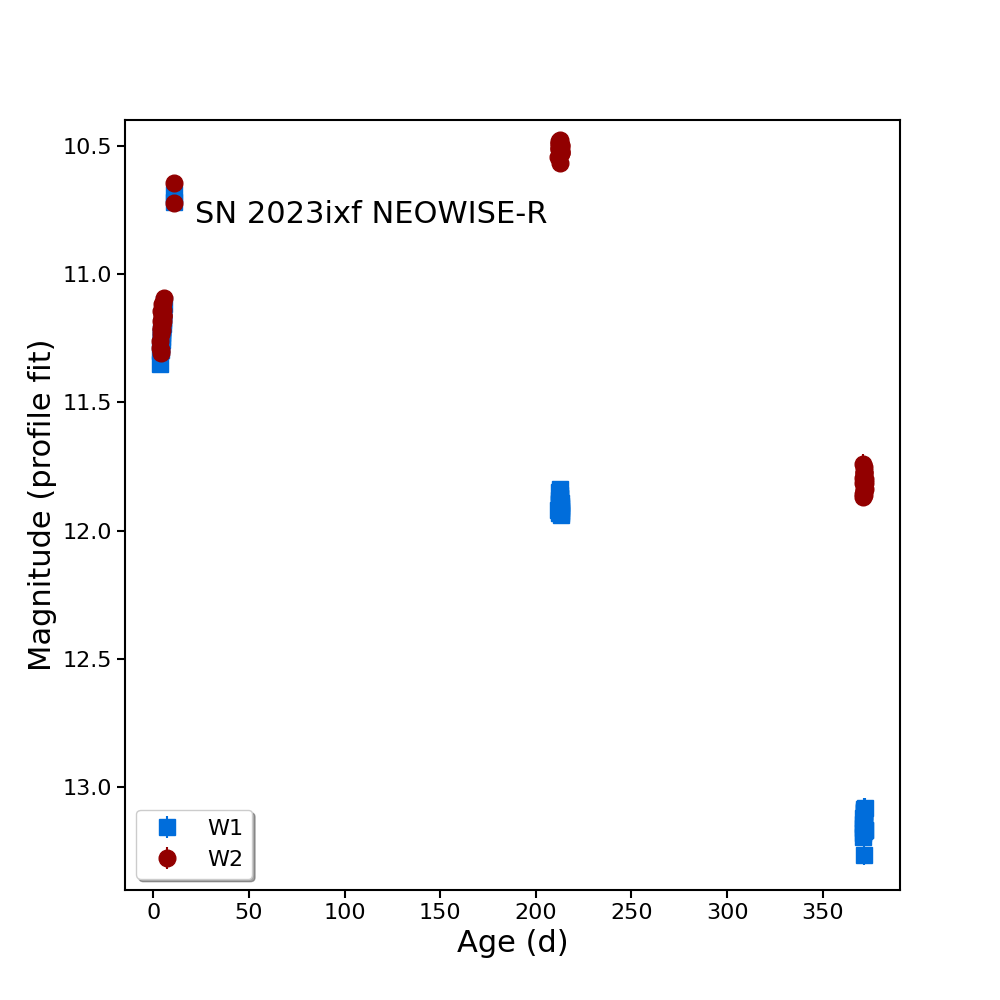}
\caption{{\sl NEOWISE-R\/} light curves of SN 2023ixf at 3.4 and 4.6 $\mu$m (bands W1 and W2, respectively). The observed magnitudes shown are in the Vega system (see Table~\ref{tab:neowise_obs}). They have not been corrected for reddening. The left panel is limited to the early times between 3.631 and 10.901 d, in which the steady rise in the SN brightness is evident. The right panel shows the entire set of detections, at both early and late times; for the latter, note the dramatic reddening in the color of the SN in the two bands at late times, relative to the early times.
\label{fig:neowise-lc}}
\end{figure*}

The SN itself was detected during the nineteenth sky coverage since the start of the Reactivation Mission, from day 3.631 through 10.901 (2023 May 22.38 through May 29.65). We note that the gap of several days in the early-time data is caused by a ``Moon toggle'' procedure, in which the spacecraft pointing skips ahead to avoid the Moon, and then slews back to observe the ``skipped'' area of sky (see \citealt{Wright2010} for a description); in our case, this was advantageous, as we were able to sample the SN's evolution about a week after the first detections. All of these data are publicly available via the NASA/IPAC Infrared Science Archive (IRSA; \href{https://irsa.ipac.caltech.edu/}{https://irsa.ipac.caltech.edu/}); see Figure~\ref{fig:neowise-image}. The SN was then captured again at late times during the twenty-first sky coverage, from day 211.736 through 213.351 (2023 December 16.49 through 18.10), and then again from day 370.875 through 372.476 (2024 May 23.63 through 25.23). Those data were still pre-release at the time of this writing and were also obtained via IRSA. Given the separation of the SN from the general environs of the neighboring giant H~{\sc ii} region NGC~5461, the detections are relatively clean. The SN detections are listed in Table~\ref{tab:neowise_obs}. The quantities {\tt W1mpro} and {\tt W2mpro} are profile-fit photometry magnitudes at 3.4 and 4.6 $\mu$m (bands W1 and W2), respectively, in the Vega system\footnote{See the NEOWISE Data Release Explanatory Supplement, \href{https://wise2.ipac.caltech.edu/docs/release/neowise/expsup/}{https://wise2.ipac.caltech.edu/docs/release/neowise/expsup/}.}, without any further special processing or additional background subtraction applied. The resulting light curves are shown in Figure~\ref{fig:neowise-lc}.

\startlongtable
\begin{deluxetable*}{cccccccc}
\tablewidth{0pt}
\tablecolumns{8}
\tablecaption{{\sl NEOWISE-R\/} Observations of SN 2023ixf \label{tab:neowise_obs}}
\tablehead{\colhead{MJD} & \colhead{Age} & \colhead{Scan} & \colhead{Frame} & \colhead{W1mpro} & \colhead{W1sigmpro} & \colhead{W2mpro} & \colhead{W2sigmpro} \\
\colhead{} & \colhead{(d)} & \colhead{ID} & \colhead{Num} & \colhead{(mag)} & \colhead{(mag)} & \colhead{(mag)} & \colhead{(mag)}}
\startdata
60086.381 &   3.631 & 50719r & 232 & 11.352 & 0.020 & 11.287 & 0.026 \\
60086.511 &   3.761 & 50723r & 159 & 11.301 & 0.022 & 11.262 & 0.036 \\
60086.641 &   3.891 & 50727r & 232 & 11.296 & 0.019 & 11.212 & 0.028 \\
60086.771 &   4.021 & 50731r & 232 & 11.241 & 0.020 & 11.309 & 0.026 \\
60086.836 &   4.086 & 50733r & 207 & 11.278 & 0.021 & 11.233 & 0.026 \\
60086.836 &   4.086 & 50733r & 208 & 11.276 & 0.023 & 11.296 & 0.027 \\
60086.901 &   4.151 & 50735r & 231 & 11.225 & 0.020 & 11.183 & 0.028 \\
60086.966 &   4.216 & 50737r & 207 & 11.247 & 0.019 & 11.220 & 0.031 \\
60087.031 &   4.281 & 50739r & 232 & 11.260 & 0.018 & 11.146 & 0.025 \\
60087.096 &   4.346 & 50741r & 208 & 11.258 & 0.020 & 11.219 & 0.028 \\
60087.160 &   4.410 & 50743r & 157 & 11.208 & 0.018 & 11.198 & 0.026 \\
60087.225 &   4.475 & 50745r & 157 & 11.245 & 0.021 & 11.156 & 0.025 \\
60087.290 &   4.540 & 50747r & 232 & 11.228 & 0.018 & 11.178 & 0.026 \\
60087.355 &   4.605 & 50749r & 208 & 11.220 & 0.020 & 11.196 & 0.026 \\
60087.420 &   4.670 & 50751r & 232 & 11.200 & 0.020 & 11.169 & 0.026 \\
60087.484 &   4.734 & 50753r & 157 & 11.354 & 0.020 & 11.451 & 0.027 \\
60087.485 &   4.735 & 50753r & 158 & 11.208 & 0.020 & 11.180 & 0.032 \\
60087.549 &   4.799 & 50755r & 156 & 11.209 & 0.017 & 11.115 & 0.023 \\
60087.615 &   4.865 & 50757r & 208 & 11.195 & 0.020 & 11.162 & 0.028 \\
60087.680 &   4.930 & 50759r & 232 & 11.181 & 0.018 & 11.139 & 0.023 \\
60087.745 &   4.995 & 50761r & 207 & 11.145 & 0.021 & 11.112 & 0.023 \\
60087.745 &   4.995 & 50761r & 208 & 11.145 & 0.021 & 11.165 & 0.026 \\
60087.874 &   5.124 & 50765r & 207 & 11.172 & 0.018 & 11.158 & 0.024 \\
60088.004 &   5.254 & 50769r & 208 & 11.147 & 0.017 & 11.183 & 0.024 \\
60088.134 &   5.384 & 50773r & 156 & 11.117 & 0.019 & 11.095 & 0.022 \\
60093.586 &  10.836 & 50941r & 211 & 10.719 & 0.019 & 10.725 & 0.021 \\
60093.651 &  10.901 & 50942r & 235 & 10.687 & 0.016 & 10.646 & 0.020 \\
60294.486 & 211.736 & 57146r & 007 & 11.921 & 0.027 & 10.544 & 0.020 \\
60294.616 & 211.866 & 57150r & 042 & 11.920 & 0.021 & 10.505 & 0.018 \\
60294.745 & 211.995 & 57154r & 042 & 11.895 & 0.021 & 10.489 & 0.019 \\
60294.874 & 212.124 & 57158r & 042 & 11.880 & 0.020 & 10.513 & 0.018 \\
60294.938 & 212.188 & 57160r & 017 & 11.849 & 0.020 & 10.513 & 0.018 \\
60295.003 & 212.253 & 57162r & 042 & 11.891 & 0.022 & 10.493 & 0.018 \\
60295.068 & 212.318 & 57164r & 018 & 11.930 & 0.023 & 10.483 & 0.019 \\
60295.197 & 212.447 & 57168r & 018 & 11.839 & 0.020 & 10.510 & 0.019 \\
60295.261 & 212.511 & 57170r & 042 & 11.874 & 0.023 & 10.569 & 0.025 \\
60295.326 & 212.576 & 57172r & 018 & 11.860 & 0.020 & 10.511 & 0.018 \\
60295.390 & 212.640 & 57174r & 042 & 11.929 & 0.024 & 10.508 & 0.021 \\
60295.455 & 212.705 & 57176r & 017 & 11.901 & 0.021 & 10.495 & 0.019 \\
60295.584 & 212.834 & 57180r & 017 & 11.872 & 0.023 & 10.479 & 0.020 \\
60295.649 & 212.899 & 57182r & 042 & 11.911 & 0.025 & 10.517 & 0.021 \\
60295.713 & 212.963 & 57184r & 018 & 11.930 & 0.023 & 10.530 & 0.019 \\
60295.778 & 213.028 & 57186r & 043 & 11.925 & 0.028 & 10.498 & 0.020 \\
60295.842 & 213.092 & 57188r & 018 & 11.905 & 0.024 & 10.527 & 0.020 \\
60295.972 & 213.222 & 57192r & 018 & 11.894 & 0.023 & 10.525 & 0.023 \\
60296.101 & 213.351 & 57196r & 018 & 11.939 & 0.024 & 10.502 & 0.020 \\
60453.625 & 370.875 & 62087r & 131 & 13.141 & 0.038 & 11.795 & 0.036 \\
60453.753 & 371.003 & 62091r & 231 & 13.120 & 0.034 & 11.815 & 0.039 \\
60453.881 & 371.131 & 62095r & 232 & 13.177 & 0.036 & 11.856 & 0.040 \\
60453.945 & 371.195 & 62101r & 206 & 13.194 & 0.035 & 11.741 & 0.039 \\
60454.009 & 371.259 & 62103r & 231 & 13.157 & 0.035 & 11.867 & 0.036 \\
60454.073 & 371.323 & 62105r & 232 & 13.098 & 0.032 & 11.769 & 0.037 \\
60454.137 & 371.387 & 62107r & 134 & 13.101 & 0.032 & 11.817 & 0.043 \\
60454.201 & 371.451 & 62109r & 135 & 13.172 & 0.038 & 11.775 & 0.038 \\
60454.265 & 371.515 & 62111r & 231 & 13.149 & 0.036 & 11.751 & 0.034 \\
60454.329 & 371.579 & 62113r & 206 & 13.265 & 0.038 & 11.836 & 0.037 \\
60454.393 & 371.643 & 62115r & 231 & 13.085 & 0.031 & 11.861 & 0.043 \\
60454.457 & 371.707 & 62117r & 205 & 13.091 & 0.049 & 11.775 & 0.032 \\
60454.458 & 371.708 & 62117r & 206 & 13.134 & 0.036 & 11.844 & 0.038 \\
60454.521 & 371.771 & 62119r & 231 & 13.118 & 0.032 & 11.816 & 0.041 \\
60454.585 & 371.835 & 62121r & 205 & 13.080 & 0.038 & 11.800 & 0.036 \\
60454.650 & 371.900 & 62123r & 231 & 13.168 & 0.032 & 11.815 & 0.033 \\
60454.714 & 371.964 & 62125r & 206 & 13.165 & 0.033 & 11.839 & 0.040 \\
60454.778 & 372.028 & 62127r & 231 & 13.056 & 0.031 & 11.789 & 0.034 \\
60454.842 & 372.092 & 62129r & 206 & 13.254 & 0.036 & 11.705 & 0.038 \\
60454.906 & 372.156 & 62131r & 231 & 13.053 & 0.031 & 11.819 & 0.031 \\
60454.970 & 372.220 & 62133r & 206 & 13.228 & 0.035 & 11.797 & 0.041 \\
60455.097 & 372.347 & 62137r & 125 & 13.120 & 0.034 & 11.768 & 0.036 \\
60455.226 & 372.476 & 62141r & 206 & 13.202 & 0.035 & 11.823 & 0.035 \\
\enddata
\tablecomments{The columns W1mpro, W1sigmpro, W2mpro, and W2sigmpro are profile-fit photometry magnitudes and their uncertainties in the Vega system.}
\end{deluxetable*}

\subsection{Late-Time Optical Data}

\subsubsection{Konkoly Observatory}

Many investigators have continued to follow SN~2023ixf since its discovery. 
We (Vink\'o, Jo\'o, P\'al, Kriskovics, K\"onyves-T\'oth, Szak\'ats, Vida) obtained optical photometry with the 0.8 m Ritchey-Chr\'etien telescope at the Konkoly Observatory, Hungary (J.~Vink\'o et al.~2024, in preparation). This includes late-time Johnson $BV$ and Sloan Digital Sky Survey (SDSS) $g'r'i'z'$ (hereafter $BVgriz$) photometry from MJD 60297.0 (2023 December 18.5, day 214) and from MJD 60444.93 (2024 May 14.9, day 362.18).
These data were processed with standard IRAF\footnote{IRAF is distributed by the National Optical Astronomy Observatories, which are operated by the Association of Universities for Research in Astronomy, Inc., under a cooperative agreement with the National Science Foundation.} routines. Photometric calibration was based on field stars in the Pan-STARRS DR1 (PS1) catalogue\footnote{\href{https://catalogs.mast.stsci.edu/panstarrs/}{https://catalogs.mast.stsci.edu/panstarrs/}} \citep{Chambers_2016}. In order to obtain reference magnitudes for our $B$ and $V$ frames, the PS1 magnitudes were transformed into the Johnson-Cousins $BVRI$ system, based on equations and coefficients provided by \cite{Tonry_12}. Finally, the instrumental magnitudes were transformed into standard $BVgriz$ magnitudes by applying a linear color term (using $g-i$) and wavelength-dependent zero points. Since the reference stars were all within a few arcminutes around the SN, no atmospheric extinction correction was necessary.

\subsubsection{Lick Observatory}

We (Zheng, Brink, Filippenko) performed further follow-up $BVRI$ photometric observations of SN~2023ixf with both the 0.76~m Katzman Automatic Imaging Telescope (KAIT) and the 1~m Nickel telescope, as well as spectroscopy with the Shane 3~m telescope at Lick Observatory; see W.~Zheng et al.~(2024, in preparation) for details of the observations and data reduction. Regarding the photometry, we isolated just the late-time data on MJD 60291.53 and 60303.59 (2023 December 13.53 and 25.59, or days 208.78 and 220.84, respectively), which bracketed the late-time {\sl NEOWISE\/} observations, and interpolated between these two sets of measurements. As far as spectroscopy, we have included spectra from Zheng et al.~obtained on 60290.53 (2023 December 12.53, day 207.78) and 60445.41 (2024 May 15.41, day 362.66), which are the closest Lick spectra in time to the late-time {\sl NEOWISE\/} observations.

\section{Analysis}

Throughout our analysis we have assumed a total extinction to SN 2023ixf of $A_V=0.12$ mag from \citet{VanDyk2024}. For UV through the near-IR we adopted the \citet{Fitzpatrick1999} reddening law. The extinction corrections for the {\sl NEOWISE\/} bands are adopted from \citet{Wang2019}.

\subsection{Early-Time IR Emission}\label{sec:early_emission}

Rather than select every pair of observed {\sl NEOWISE\/} W1 and W2 data points to analyze, for illustrative purposes we have chosen to consider just two sets at early times, the very first one from MJD 60086.381 (day 3.631) and from 60093.586 (day 10.836). These adequately represent the two periods of early-time sampling of the light curves in these bands.

In order to put the {\sl NEOWISE\/} data in context with the overall SED at these two epochs, we accumulated published light-curve data at UV, optical, and near-IR wavelengths corresponding to (or bracketing) the epochs. The data sources were then an amalgam of {\sl Swift\/} UVW2, UVM2, UVW1 from \citet{Zimmerman2024}, SDSS $ugriz$ and Johnson $BVJHK_s$ from \citet{Teja2023}, Johnson $UBV$, and SDSS $griz$ from \citet{Hiramatsu2023}, and Johnson $JHK$ from \citet{Yamanaka2023}. 

Since the {\sl NEOWISE\/} measurements are in the Vega system, the entirety of the dataset presented by \citet{Teja2023}, which is in the AB system, had to be converted to Vega magnitudes. The SDSS magnitudes from \citet{Hiramatsu2023} also required a similar conversion. No conversion was needed for the \citet{Yamanaka2023} $JHK$ photometry. For day 3.631 the available UV-optical data were quasi-contemporaneous with the {\sl NEOWISE\/} points; however, $JHK$ required a linear interpolation between two bracketing epochs (the earlier epoch was at day 3.4, very close in time to {\sl NEOWISE}). For day 10.836 none of the complementary data were contemporaneous, so we were forced to interpolate between bracketing epochs at all wavelengths.

\begin{figure*}
\centering
\includegraphics[width=.48\textwidth]{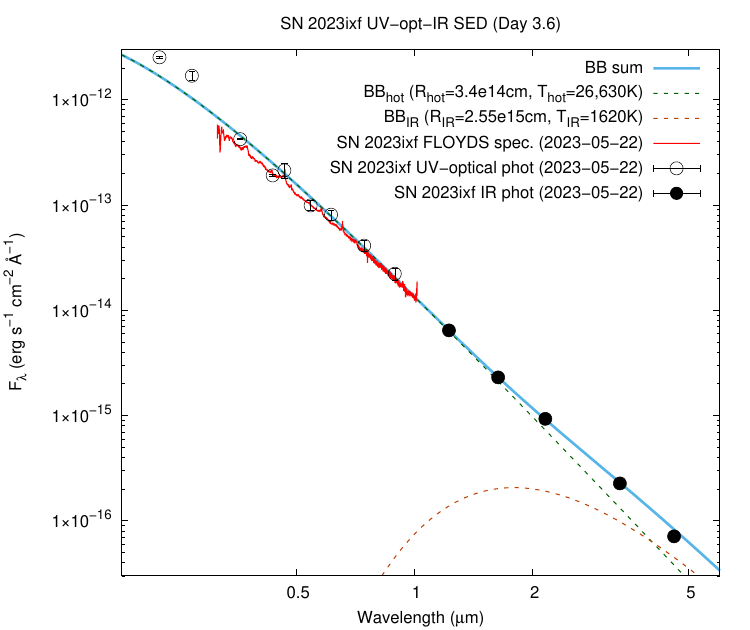}
\includegraphics[width=.48\textwidth]{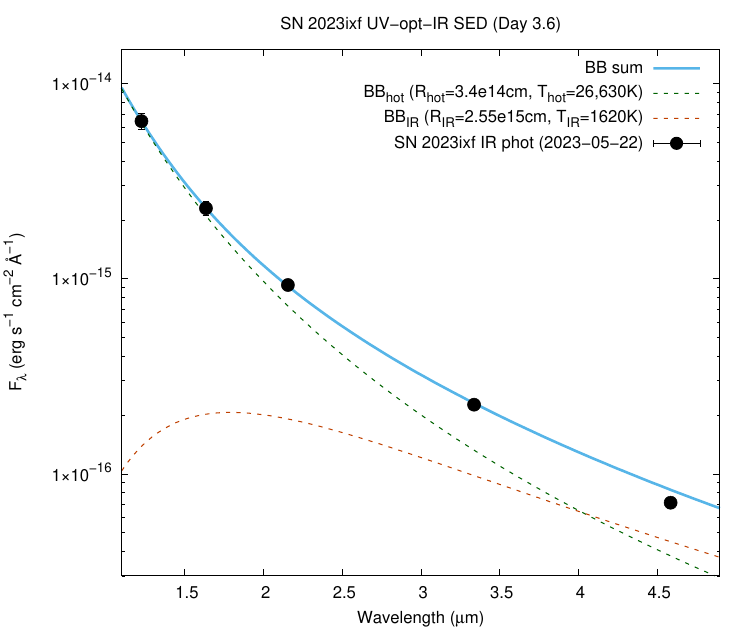}
\caption{{\it Left:} The combined, dereddened UV-optical-IR SED of SN~2023ixf on day 3.6; see text for sources of the photometry. For comparison, we also show a dereddened, renormalized FTN/FLOYDS spectrum from day 3.54 \citep[][ solid red curve]{Bostroem2023}. Blackbody SED fitting of only the near-IR and mid-IR fluxes (filled circles) required two components: One component consisting of a hot ($\sim 26,630$~K) blackbody (dashed green curve), and the other a cooler ($\sim 1,620$~K) blackbody (``IR''; dashed orange curve). The total fit is shown as the solid blue curve. The inferred radii of the blackbodies are $R_{\rm hot} \approx 3.4 \times 10^{14}$ and $R_{\rm IR} \approx 2.6 \times 10^{15}$ cm. {\it Right:} A zoom-in of just the IR portion of the SED.}
\label{fig:SED_3_6}
\end{figure*}

The resulting SED is shown in Figure~\ref{fig:SED_3_6}. In addition to the UV-optical photometric points we included an FTN-FLOYDS-N spectrum of the SN obtained by \citet{Bostroem2023} on 2023 May 22, which we downloaded from WISeREP\footnote{\href{https://www.wiserep.org}{https://www.wiserep.org}} \citep{Yaron2012}. Both the spectrum and the photometry were first reddening-corrected. This spectrum was further renormalized to the (dereddened) SN $V$-band brightness. As can be seen in the figure, the overall agreement is reasonable between the spectrum and the photometric points across the common wavelength range.

We then attempted to fit a single, simple blackbody to the SED at day 3.631. We found that a hot, $\sim 26,630$~K blackbody provides a good fit to the optical data, although a clear excess exists in the UV relative to this fit. The fit implies that the SN luminosity at that epoch was $\gtrsim 4.1 \times 10^{43}$ erg s$^{-1}$; this is a lower limit, since there is clearly additional luminosity in the UV. This is consistent with the evidence for strong, early-time interaction of the SN shock with dense CSM \citep[e.g.,][]{Bostroem2023,Jacobson-Galan2023,Teja2023,Martinez2024,Zimmerman2024}; specifically, interaction in SNe II can strengthen the continuum flux and boost emission lines in the UV simultaneously \citep{Dessart_2022}. The blackbody radius is then $R_{\rm hot} \approx 3.4 \times 10^{14}$ cm. The early-phase photometric and spectroscopic UV-optical SN observations provided evidence for strong interaction of the SN shock with a dense, confined ($<2 \times 10^{15}$~cm) CSM \citep[e.g.,][]{Jacobson-Galan2023,Smith2023,Bostroem2023,Teja2023}. \citet{Zimmerman2024} further refined the extent of the dense CSM to $R_{\rm CSM} \approx 2 \times 10^{14}$ cm and concluded that it actually delayed shock breakout (SBO) from hours after explosion to $\sim 3$ d. In other words, the very first {\sl NEOWISE\/} data were likely obtained within just hours after SBO, and the inferred $R_{\rm hot}$ is consistent with the shock having already overrun the confinements of the dense CSM. In fact, following \citet{Zimmerman2024} and assuming a SN expansion velocity $v_{\rm exp}=8,000$ km s$^{-1}$, on day 3.631 the shock would have been at $\sim 2.5 \times 10^{14}$ cm, roughly consistent with $R_{\rm hot}$ (if $v_{\rm exp}$ instead had been a somewhat higher, $\sim 11,000$ km s$^{-1}$, the two radii would be in better agreement).

Particularly fascinating here is that an obvious excess in flux, relative to the hot blackbody, can be seen in Figure~\ref{fig:SED_3_6} at $\gtrsim 1.5$ $\mu$m. We found that we could account for this IR excess with an additional much cooler blackbody, at $\sim 1620$~K. Including this extra blackbody provides a reasonable fit to $JHK_s$ and W1, although it does not quite fit W2 as well. This additional source of IR emission is of a comparatively far smaller luminosity, $\sim 3.2 \times 10^{40}$ erg s$^{-1}$, than the SN shock (it accounts for $\lesssim 0.1$\% of the total emission). The corresponding blackbody radius is $R_{\rm IR} \approx 2.6 \times 10^{15}$ cm. \citet{Kilpatricketal2023} and \citet{VanDyk2024}, for example, inferred that the RSG progenitor candidate was surrounded by a dusty shell with an inner radius of $R_{\rm in} \approx (0.5$--$1.0) \times 10^{15}$ cm. The assumption \citet{VanDyk2024} made in their modeling of the star was that the shell extended out to $1000\ \times R_{\rm in}$, with the dust density declining $\propto r^{-2}$. Furthermore, $\sim 1620$~K is roughly the estimated evaporation temperature of $\sim 0.01$--0.1 $\mu$m-sized silicate-dominated dust in SN environments \citep[e.g.,][]{gall14}. (Note that \citealt{VanDyk2024} were able to fit the reddening-corrected SED of the pre-explosion dust shell with a simple $\sim 1761$~K blackbody.) Thus, we speculate that the IR excess was emanating from the dusty CSM shell, with the SN shock still within $R_{\rm in}$. 

This analysis including the {\sl NEOWISE\/} data lends credence to the overall picture of the progenitor star, inferred via the modeling of the observed SED of the star. The estimated luminosity from the optically-thin dust was still about two orders-of-magnitude larger than the luminosity of the progenitor candidate ($\sim 9 \times 10^4\ L_{\odot}$, or $\sim 3.5 \times 10^{38}$ erg s$^{-1}$; \citealt{VanDyk2024}), which implies that the dust shell was likely heated by and was reprocessing the UV emission from the interaction of the SN shock with the dense inner CSM. We note that much of the CSM dust was likely destroyed immediately after explosion by high-energy (extreme UV to $\gamma$-ray) photons from the blast, through grain sublimation, vaporization, and extreme grain charging effects \citep{Jones2004}.

\bibpunct[;]{(}{)}{;}{a}{}{;}

\begin{figure*}
\centering
\includegraphics[width=.48\textwidth]{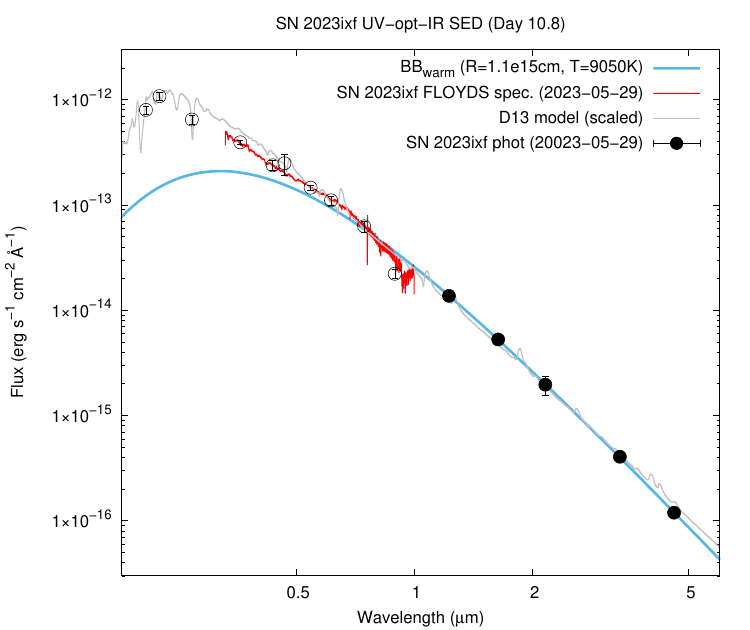}
\includegraphics[width=.48\textwidth]{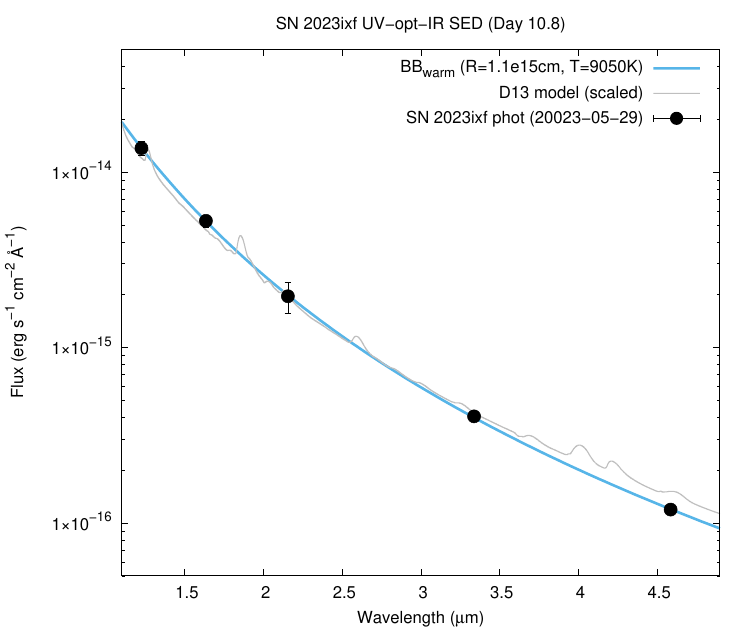}
\caption{{\it Left:} The combined, dereddened UV-optical-IR SED of SN~2023ixf on day 10.8; see text for sources of the photometry. For comparison, we also show a dereddened, renormalized FTN/FLOYDS spectrum from day 10.73 \citep[][red curve]{Bostroem2023}. Blackbody fitting of the SED was performed, consisting of only one component (solid blue curve) at $\sim 9,050$~K, with only the near-IR and mid-IR fluxes (filled circles) used in the fitting. The fit is poor for wavelengths $< 1\ \mu$m. The addition of a scaled version of the {\it m15mlt1} model spectrum at day 11 (solid gray curve) from \citet[][ D13]{dessart13} provides a much better representation of the observed data. The inferred radius of the blackbody is $R_{\rm warm} \approx 1.1 \times 10^{15}$ cm. {\it Right:} A zoom-in of just the IR portion of the SED. Note that the D13 model provides an overall reasonable representation of the SED, even for the {\sl NEOWISE\/} bands.}
\label{fig:SED_10_8}
\end{figure*}

We built a similar SED from the available UV/optical/near-IR/{\sl NEOWISE\/} photometry from day 10.836. Again, we added to this a dereddened and renormalized FTN-FLOYDS-N SN spectrum from WISEReP from 2023 May 29 \citep{Bostroem2023}. The resulting SED is shown in Figure~\ref{fig:SED_10_8}. We attempted to fit a warm, $\sim 9050$~K blackbody to the data; however, as can be seen in the figure, while this fit is very good in the IR, it diverges significantly for wavelengths $< 1\ \mu$m. If we consider the radiation models for SNe~IIP by \citet[][ D13]{dessart13}, specifically, the {\it m15mlt1\/} model (with an initial progenitor mass of $15\ M_{\odot}$ and radius $R_{\star}=1107\ R_{\odot}$) computed at day 11, it provides a remarkably good comparison, from the UV through to the mid-IR, with the observed data. This implies that what we were seeing at that epoch, including with {\sl NEOWISE}, was SN ejecta-dominated emission. If we take into account the blackbody fit, the inferred luminosity is $\gtrsim 5.8 \times 10^{42}$ erg s$^{-1}$, which is a lower limit, since a significant amount of luminosity is still emerging from the SN at $< 1\ \mu$m. In fact, if we integrate the D13 model (for instance), we obtain a luminosity of $\sim 1.5 \times 10^{43}$ erg s$^{-1}$. The blackbody radius is $R_{\rm warm} \approx 1.1 \times 10^{15}$ cm. Once again, assuming $v_{\rm exp}=8,000$ km s$^{-1}$, the shock radius would have been $\sim 7.5 \times 10^{14}$ cm, roughly consistent with $R_{\rm warm}$; similar to day 3.631, a somewhat higher $v_{\rm exp} = 11,000$ km s$^{-1}$ would result in better agreement between the two radius estimates.

The blackbody radii and temperatures that we have calculated for both days 3.6 and 10.8 are in a good agreement with the results of \citet{Singh_2024}. Note, however, that we have assumed a spherical and homogeneous medium, while early-phase spectropolarimetric data implied the presence of an aspherical dense CSM and a clumpy, low-density extended CSM around SN~2023ixf \citep{Vasylyev_2023,Singh_2024}. Nevertheless, our conclusions also fit with that of \citet{Vasylyev_2023}, in that they concluded that the expanding SN ejecta just emerged from the dense CSM region around Day 3.5.

\subsection{Late-Time Mid-IR Excess}

We computed the averages of the various measurements obtained by {\sl NEOWISE\/} at late times in the SN's evolution, between MJD 60294.5 and 60296.1. SN 2023ixf was already well on the radioactively-powered exponential tail by this point. The averages are $11.895 \pm 0.023$ mag and $10.509 \pm 0.020$ mag in W1 and W2, respectively, with reference to day 213.

\begin{figure*}
\centering
\includegraphics[width=.65\textwidth]{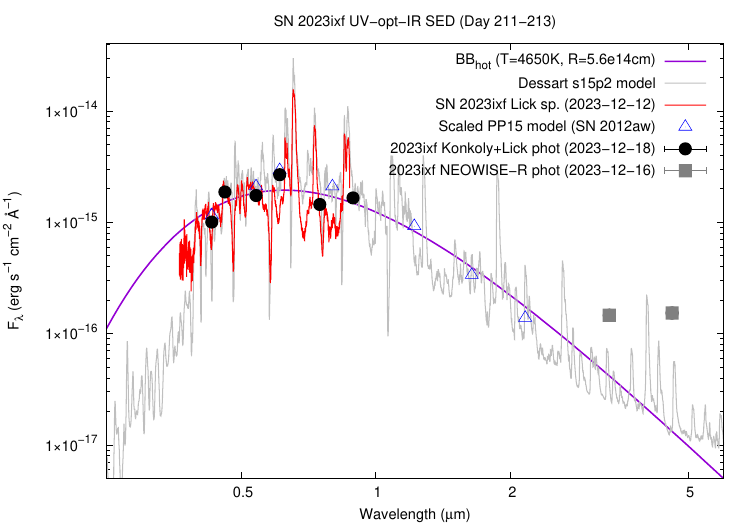}
\caption{One-component blackbody (BB) fit on the Days 211--213 optical SED of SN~2023ixf (filled circles). {\sl NEOWISE-R\/} data (filled gray squares) were excluded from the fit. Atmospheric models for SNe IIP are also shown for comparison: The day 207 {\it s15p2\/} model spectrum (courtesy of Luc Dessart, gray curve), and the distance-scaled model fluxes interpolated between days 193 and 242, for SN~2012aw from \citet[][ open blue triangles]{PP2015}; the distance to SN~2012aw ($D=9.9$ Mpc) was adopted from \citet{Bose_2013}.}
\label{fig:SED_213_BB_IR}
\end{figure*}

Thanks to the active worldwide follow-up observations of SN 2023ixf, we can also build the late-time optical-NIR SED for the SN. We assembled the unpublished optical $BVgriz$ photometry from Konkoly Observatory (from 2023 December 18.5, day 214; J.~Vink\'o et al. 2024, in preparation) and Lick Observatory $BVRI$ data (interpolated to the same date), as well as a single Lick optical spectrum (from 2023 December 12, day 208; W. Zheng et al. 2024, in preparation); see Section \ref{sec:observations}. We show the combined optical SED in Figure~\ref{fig:SED_213_BB_IR}. Since we are unaware of any available near-IR data close to this late epoch, we included spectra and fluxes from models of SN~IIP explosions for comparison: The day 207 {\it s15p2} model spectrum courtesy of Luc Dessart, and the interpolated and distance-scaled day +193 and +242 model fluxes for the Type IIP SN~2012aw from \citet{PP2015}. (We compared the Fe~{\sc ii} line velocities for SN 2023ixf from Zheng et al.~2024, to that of SN~2012aw from \citealt{Bose_2013} and found a very good match even at late phases; we also adopted the SN~2012aw distance of 9.9 Mpc from \citealt{Bose_2013}.) Both the model spectrum and the set of optical model fluxes appear to compare well with our late-time measured optical data; thus, we find it to be a reasonable approach to use the near-IR components of these models as references during the analysis of the late-time optical-IR SED.

As can be seen in Figure~\ref{fig:SED_213_BB_IR}, the {\sl NEOWISE\/} 3.4 and 4.6 $\mu$m data show a clear excess relative to the model fluxes. We fit a single blackbody to the optical-near-IR part of the SED and found a warm component with temperature $\sim 4650$ K and corresponding radius $\sim 5.6 \times 10^{14}$ cm. While SN ejecta are not expected to be realistically represented by a blackbody at late epochs, we can then use the result of this fit to provide a characterization of the colder, longer-wavelength excess that is apparent.

As we noted in Section~\ref{sec:early_emission}, we assume that the majority of pre-existing dust grains, located in the dense, confined circumstellar shell, have been evaporated at early times after explosion; therefore, these can no longer be responsible for the excess. Thus, we consider two possible alternate scenarios: (i) At day 213, newly-formed dust existed either in the inner (unshocked) ejecta or in the contact discontinuity between the forward and reverse shocks (the CDS); or (ii) radiation was being emitted by more distant pre-existing dust grains in the SN environment, heated collisionally or radiatively by the (forward) SN shock. However, from only the two mid-IR data points, a detailed investigation of the origin and properties of the assumed late-time dust is challenging to carry out --- we hope that pending {\sl JWST\/} observations of the SN will provide greater insight into these two scenarios. Furthermore, we note that the large 4.6 $\mu$m flux excess at day 213 can also be explained by the emergence of the 1--0 vibrational band of carbon monoxide (CO) at 4.65 $\mu$m, as seen in some SNe IIP with observed mid-IR spectra at a similar age \citep[e.g.,][]{kotak05,kotak06,szalai11,szalai13}, and also predicted by modeling of exploding RSGs by \citet{McLeod_2024}. The presence of a late-time mid-IR excess is in agreement with the results from \cite{Singh_2024}; they found that the flattening in the $K_s$-band light curve and the attenuation of the red-edge of the H$\alpha$ line profile $\sim$125 d after explosion is indicative of the onset of molecular CO and, hence, dust formation in SN~2023ixf. This possibility should be also taken into account during any analysis.

\begin{figure*}
\centering
\includegraphics[width=.65\textwidth]{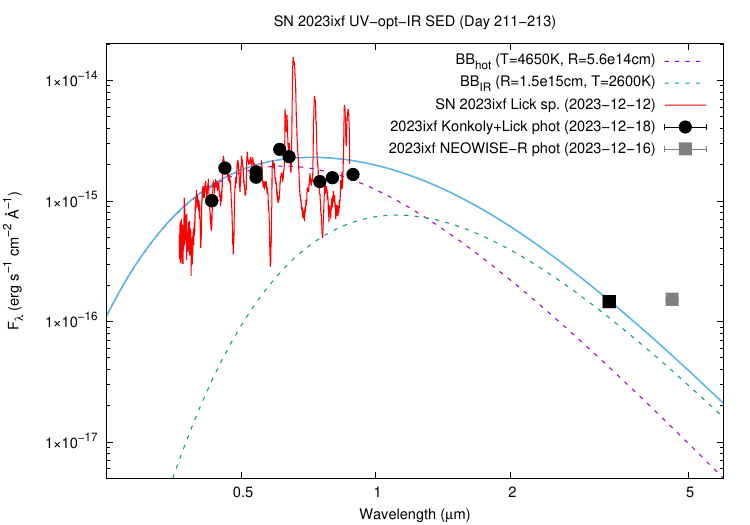}
\includegraphics[width=.65\textwidth]{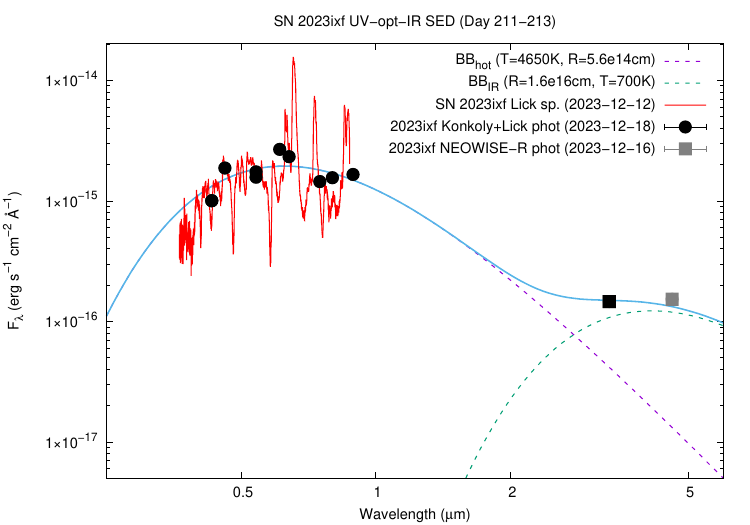}
\caption{The same as in Fig. \ref{fig:SED_213_BB_IR} but with two-component BB fits on the combined Days 211--213 optical-IR SED of SN~2023ixf, and without showing atmospheric models. The {\sl NEOWISE-R\/} W2 point was excluded from the fit because of the potential contamination from CO line emission. Parameters of the hot component are the same as above. The panels show two scenarios for the ``warm'' dust component: one calculated assuming the highest theoretical dust temperature ($T_\textrm{IR} = 2600$~K, top panel), and another one assuming $T_\textrm{IR} = 700$~K (bottom panel); see details in the text.}
\label{fig:SED_213_BB_IR2}
\end{figure*}

Thus, we also excluded the W2 flux and fit a two-component blackbody model to the remainder of the optical-IR SED, holding the parameters from the hot component fixed. Since a fit to only one mid-IR point would not provide any physically relevant information, we applied temperature constraints and assumed two possible scenarios for the ``warm'' dust component: (i) First, we assumed the highest theoretical dust temperature possible for amorphous carbon dust, $T_{\rm IR} = 2600$~K (see, e.g., \citealt{gall14}); and, (ii) second, we assumed a ``typical'' temperature of ``warm'' dust ($T_{\rm IR} = 700$~K) seen in SNe~IIP at a similar age \citep[see, e.g.,][]{kotak09,szalai11,szalai13}. Using these two assumptions, we found a blackbody radius of $R_{\rm IR} \sim 1.5 \times 10^{15}$ cm and $\sim 1.6 \times 10^{16}$ cm, respectively, for the two assumptions; see Figure~\ref{fig:SED_213_BB_IR2}).

Extrapolating the Fe~{\sc ii} velocities (W.~Zheng et al. 2024, in preparation) to day 213, we estimated $\sim 1400$--1500 km s$^{-1}$ for the ejecta velocity, which results in (2.6--2.8) $\times 10^{15}$ cm for the ejecta radius. This implies that, in the case of very high-temperature dust (the first scenario above), these grains possibly could be within the ejecta. However, assuming a more realistic dust temperature (the second scenario), the dust would more likely be outside the ejecta. As \citet{VanDyk2024} concluded, the dusty shell of the progenitor likely had an inner radius of $\sim 10^{15}$ cm, with an $r^{-2}$ density distribution extending outward. Thus, the pre-explosion shell itself could easily have extended out to $\sim 1.6 \times 10^{16}$ cm (well beyond the confined volume); the dust we infer here could have been located within the CSM, being either pre-existing or newly-formed in the CDS.

\begin{figure*}
\centering
\includegraphics[width=.65\textwidth]{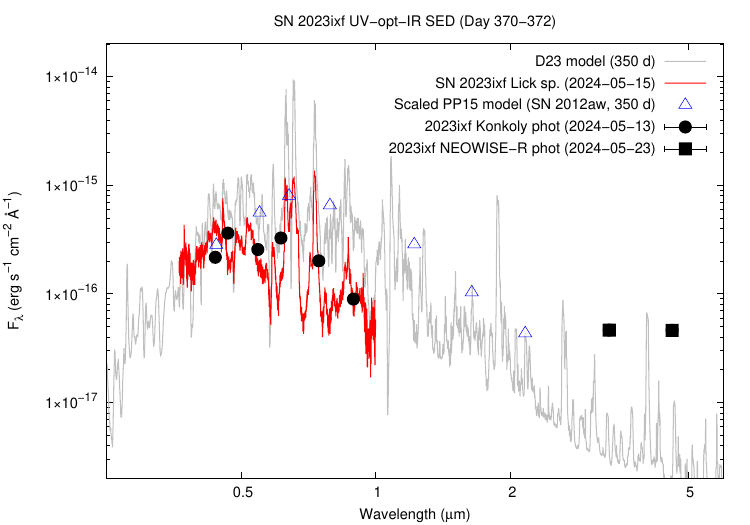}
\caption{Optical-IR SED of SN~2023ixf from days 370--372. Atmospheric models for SNe IIP are also shown for comparison: The day 350 {\it s15p2\/} model spectrum from \citet[][ D23]{Dessart_2023} and the distance-scaled PP15 models for the same epoch.}
\label{fig:SED_360}
\end{figure*}

We present in Figure~\ref{fig:SED_360} an optical-IR SED comprised of data obtained in the day 370--372 interval. Since at that range of epochs the shape of the measured optical SED appears to differ significantly from Type IIP atmospheric models by either \citet{Dessart_2023} or \citet{PP2015}, we are unable to directly estimate the amount of IR excess in the manner that we did for the day 211--213 SED. Nevertheless, clear excesses at both 3.3 and 4.6 $\mu$m are obvious.

\subsection{Comparison of SN 2023ixf with Other SNe IIP In the Mid-IR}

\begin{figure*}
\centering
\includegraphics[width=.48\textwidth]{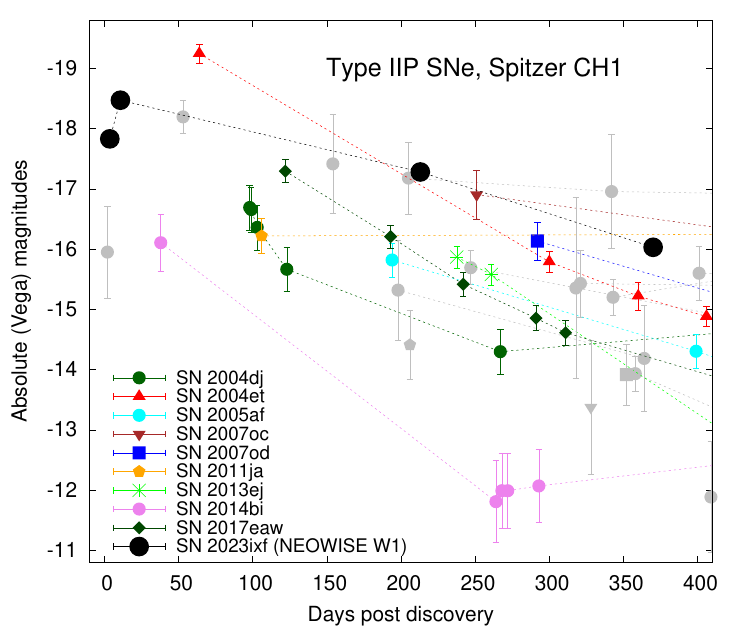}
\includegraphics[width=.48\textwidth]{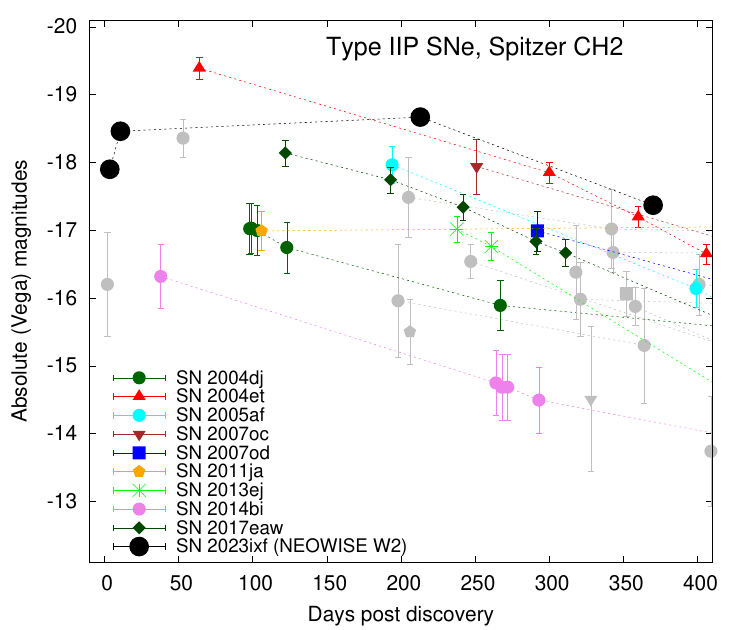}
\caption{Mid-IR luminosity evolution of SN~2023ixf ({\sl NEOWISE\/} W1 and W2 data, filled black circles), compared with {\sl Spitzer\/} IRAC data for other SNe IIP \citep{szalai19a}. Well-sampled objects are highlighted with colored symbols, while all other published detections are marked with gray symbols and are not listed in the figure legend. The earliest gray symbol shown is SN 2009js, at 2 days post-explosion \citep{Gandhi_2013,szalai19a}.}
\label{fig:midIR-LC}
\end{figure*}

We have considered all of the {\sl NEOWISE\/} data for SN 2023ixf, in the context of the mid-IR emission from other SNe, particularly SNe~IIP. To achieve this, we compared SN 2023ixf with the sample of SNe from \citet{szalai19a}, who presented photometric data obtained by {\sl Spitzer\/} in the IR Array Camera \citep[IRAC;][]{Fazio2004} bands at 3.6 and 4.5 $\mu$m during both the cryogenic and Warm missions; see Figure \ref{fig:midIR-LC}. While the bandpasses for {\sl Spitzer\/} IRAC  differ slightly from those of {\sl NEOWISE\/} W1 and W2 channels, we can draw some basic conclusions. 

First, as noted above, SN~2023ixf has become {\it the best-observed\/} SN IIP in the mid-IR during the first several days after explosion. Second, SN~2023ixf is {\it one of the most luminous\/} SNe IIP in the mid-IR ever seen. This statement is especially striking, considering the 4.5/4.6 $\mu$m photometric evolution of SNe IIP (Figure \ref{fig:midIR-LC}; right panel) --- at day 213, SN~2023ixf is more luminous than at early times and than any other SNe IIP detected so far at late times. We can speculate on why this is the case. The early-time mid-IR luminosity is consistent with the overall excess observed at other wavelengths and can be explained by the shock-CSM interaction. Among the {\sl Spitzer\/} sample, most of the SNe were either of low luminosity (e.g., SN 2004dj) or were otherwise normal (e.g., SN 2007od, SN 2011ja); only SN 2004et was observed to be somewhat extraordinary \citep[e.g.,][ see also \citealt{Shahbandeh_2023}]{Maguire2010}. The late-time mid-IR excess for SN 2023ixf could be due to post-explosion dust formation, as we have mentioned above. However, we cannot say much more about this, based on the {\sl NEOWISE\/} data alone; observations with {\sl JWST\/} will likely provide significantly more insight on this.

\section{Conclusions}

We have analyzed serendipitous observations of SN 2023ixf made as part of routine {\sl NEOWISE\/} survey scanning operations, starting on day 3.6 through day 10.9 after explosion, and again at late times from days 211 through 213 and days 370 through 372. For the three epochs in these time ranges that we analyzed, we combined the {\sl NEOWISE\/} observations with data from the UV through the near-IR, whenever possible. At day 3.6 we approximated the emission in the optical with a hot, $\sim 26,630$~K blackbody, exhibiting a marked excess in the UV, likely resulting from strong, early SN shock-CSM interaction. In the IR, however, a definite excess is also obvious, and we fit that with a cooler, $\sim 1,620$~K blackbody, with a radius of $\sim 2.6 \times 10^{15}$ cm. We concluded that this is consistent with dust in an inferred circumstellar shell surrounding the progenitor star having been heated by the UV emission from the early CSM interaction. On day 10.8 the emission, including  that detected with {\sl NEOWISE}, was consistent with being SN ejecta-dominated. 

At late times we also observed an obvious excess in the {\sl NEOWISE\/} bands, relative to the other wavelengths. This excess could arise either from newly-formed dust in the inner ejecta or in the contact discontinuity between the forward and reverse shocks (the CDS), or from more distant pre-existing dust grains in the SN environment. Furthermore, the observed large excess at 4.6 $\mu$m at late times can also be explained by the emergence of the CO 1--0 vibrational band, seen in other SNe IIP. Observations with {\sl JWST\/} are necessary to confirm detection of the CO band, as well as to better explore the overall nature of the late-time mid-IR emission.

We found, from comparing to mid-IR data for other SNe IIP, that SN 2023ixf is the best-observed SN IIP in the mid-IR during the first several days after explosion and one of the most luminous SNe IIP ever seen in the mid-IR. The survey operations by the {\sl WISE} mission, in all its incarnations, are scheduled to be terminated permanently on 2024 July 31. Together with the decommissioning of {\sl Spitzer\/} over four years ago, the number of available facilities to gather mid-IR light from nearby SNe will be greatly diminished. The next avenue will be provided by {\sl NEO Surveyor}, set for launch no later than mid-2028 \citep{Mainzer2023}: The mission will be obtaining four detections at 4.6 and 8 $\mu$m over a six-hour time period, approximately every 13 days, as part of survey operations, so it may be possible once again to catch SNe both on the rise and at late times. For future pointed observations it falls on {\sl JWST\/} to be the platform for observing both new and old SNe, to explore further in detail the nature of dust associated with these spectacular events.

\bigskip

We thank Luc Dessart for providing the day 207 model SN IIP spectrum.
This publication makes use of data products from the {\sl Near-Earth Object Wide-field Infrared Survey Explorer}, which is a joint project of the Jet Propulsion Laboratory/California Institute of Technology and the University of Arizona. 
It also uses data products from the {\sl Wide-field Infrared Survey Explorer}, which is a joint project of the University of California, Los Angeles, and the Jet Propulsion Laboratory/California Institute of Technology.  {\sl WISE\/} and {\sl NEOWISE\/} are funded by the National Aeronautics and Space Administration (NASA).
This work has been supported by the GINOP-2-3-2-15-2016-00033 project of the National Research, Development and Innovation (NRDI) Office of Hungary funded by the European Union, as well as by NKFIH OTKA FK-134432, KKP-143986, and K-142534 grants, and from the HUN-REN Hungarian Research Network. 
L.K.~and K.V.~are supported by the Bolyai J\'anos Research Scholarship
of the Hungarian Academy of Sciences. LK acknowledges the Hungarian National Research, Development and Innovation Office
grant OTKA PD-134784.
The authors acknowledge financial support of the Austrian-Hungarian Action Foundation grants 98\"ou5, 101\"ou13, 112\"ou1.
A.V.F.'s research group at UC Berkeley acknowledges financial assistance from the Christopher R. Redlich Fund, Gary and Cynthia Bengier, Clark and Sharon Winslow, Alan Eustace (W.Z. is a Bengier-Winslow-Eustace Specialist in Astronomy), William Draper, Timothy and Melissa Draper, Briggs and Kathleen Wood, Sanford Robertson (T.G.B. is a Draper-Wood-Robertson Specialist in Astronomy), and numerous other donors.   
KAIT and its ongoing operation at Lick Observatory were made possible by donations from Sun Microsystems, Inc., the Hewlett- Packard Company, AutoScope Corporation, Lick Observatory, the U.S. NSF, the    
University of California, the Sylvia \& Jim Katzman Foundation, and the TABASGO Foundation.
A major upgrade of the Kast spectrograph on the Shane 3 m telescope at Lick Observatory, led by Brad Holden, was made possible through generous gifts from the Heising-Simons Foundation, William and Marina Kast, and the University of California Observatories. 
Several UC Berkeley undergraduate students helped obtain the 1~m Nickel data.
We appreciate the excellent assistance of the staff at Lick Observatory. Research at Lick Observatory is partially supported by a generous gift from Google.

\vspace{5mm}
\facilities{NEOWISE, KAIT, Nickel, Shane, IRSA, RC80 (Konkoly)}

\software{IRAF \citep{Tody1986,Tody1993},
          PyRAF (http://www.stsci.edu/institute/\\
          software{\textunderscore}hardware/pyraf)
          }

\appendix

\section{Pre-explosion {\sl NEOWISE\/} Non-detections}\label{sec:appendix}

\begin{figure}[ht!]
\plotone{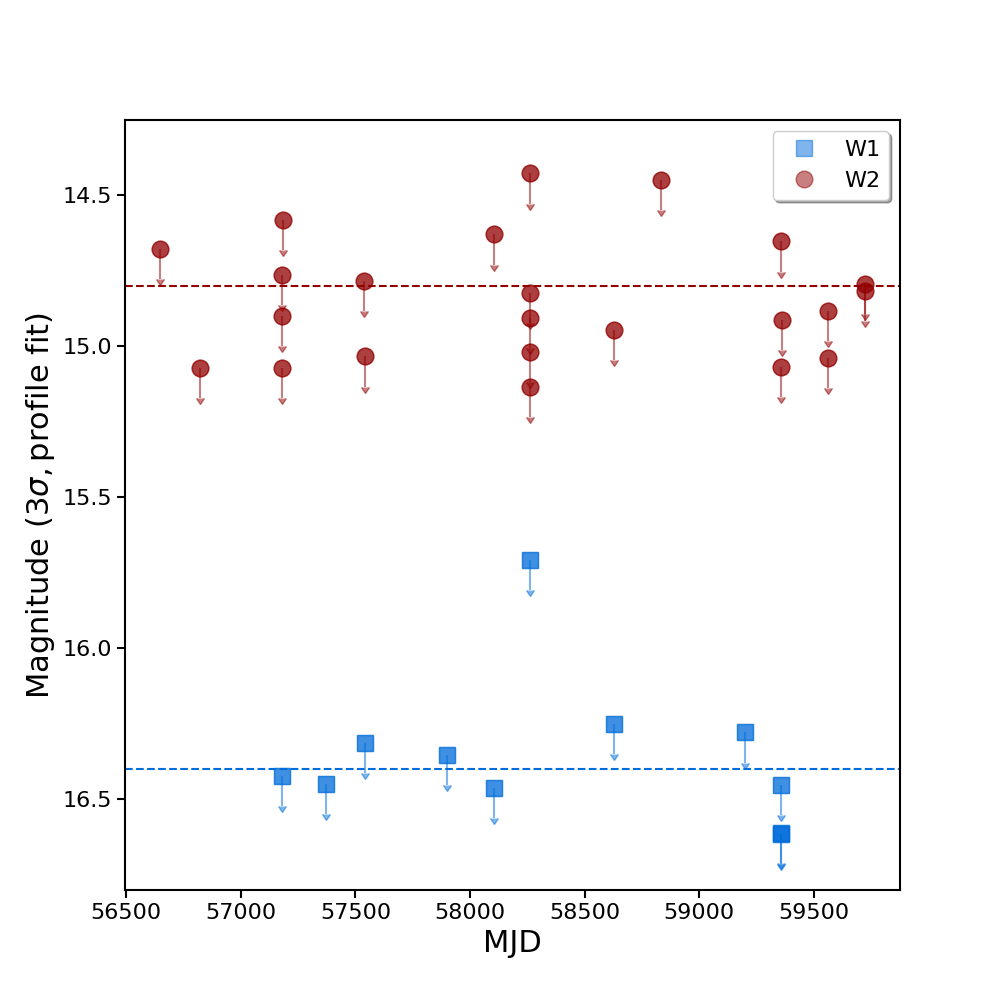}
\caption{{\sl NEOWISE-R\/} 3$\sigma$ upper limits on the detection of the SN 2023ixf progenitor candidate from between 2013 December and 2022 December at 3.4 and 4.6 $\mu$m (bands W1 and W2, respectively). Also shown are the mean values for the data at both bands (dashed lines). The observed magnitudes shown are in the Vega system. These have not been reddening-corrected.
\label{fig:upperlimits}}
\end{figure}

As pointed out in Section~\ref{sec:neowise}, {\sl NEOWISE\/} obtained pre-explosion observations of the SN site between 2013 December 18 and 2022 December 18, with the last pair of single exposures occurring 150.75 d prior to explosion. The progenitor candidate was not detected in any of these exposures. As \citet{VanDyk2024} described, the upper limits on detection were established by isolating in the {\sl NEOWISE-R\/} Single Exposure Source Table, obtained from IRSA, all of the 3$\sigma$ detected objects within 60\arcsec\ of the SN position for each band. We show these upper limits in Figure~\ref{fig:upperlimits}. Both the mean and median values in W1 are $< 16.4$ mag; the mean value in W2 is $< 14.8$ mag, whereas the median value is $< 14.9$ mag.

\bibliography{main}{}
\bibliographystyle{aasjournal}

\end{document}